\documentstyle[12pt,aaspp4]{article}

\lefthead{Hickson and Richardson}
\righthead{A curvature-compensated corrector}

\begin{document}

\title{A Curvature-Compensated Corrector for Drift-Scan Observations}
	
\author{Paul Hickson}
\affil{Department of Physics and Astronomy, University of British 
	Columbia, 2219 Main Mall, Vancouver, BC V6T 1Z4, Canada;
	paul@astro.ubc.ca}
	 
\and

\author{E. Harvey Richardson}
\affil{Dept of Mechanical Engineering, University of Victoria, Victoria, 
BC V8W 3P6, Canada; harveyr@me.uvic.ca}

\begin{abstract}

Images obtained by drift-scanning with a stationary telescope are affected by
the declination-dependent curvature of star trails.  The image displacement to
curvature and drift rate variation increases with the angular field of view and
can lead to significant loss of resolution with modern large-format CCD arrays.
We show that these effects can be essentially eliminated by means of an optical
corrector design in which individual lenses are tilted and decentered.  A
specific example is presented, of a four-element corrector designed for the
Large-Zenith Telescope.  The design reduces curvature errors to less
than $0.074\arcsec$ over a $10\arcmin \times 20\arcmin$ field of view centered
at $49\arcdeg$ declination.  By changing the positions and tilts of the lenses,
the same design can also be used for any field centers between $0\arcdeg$ and
$\pm 49\arcdeg$.

\end{abstract}

\keywords{telescopes, instrumentation: miscellaneous, surveys}
	
\section{Introduction}

In recent years the technique of time-delay integration (TDI, also known as
drift-scanning) has become well established in astronomy (\cite{MAS80},
\cite{HM84}).  For this technique, the telescope remains stationary, or tracks
at a non-sidereal rate, resulting in a steady drift on images across a CCD
detector (which is aligned so that the drift direction coincides with the
orientation of the CCD columns).  The CCD is scanned continuously at a rate
which moves the photon-produced charge along the columns at the same speed as
the optical image.  This prevents charge spread and results in sharp images
with an integration time equal to the drift-time of the image across the
CCD.

TDI offers several benefits, compared to conventional imaging.  A chief
advantage is that variations in pixel-to-pixel sensitivity are greatly reduced
because response is averaged over all CCD pixels in each column.  The required
flat field correction is essentially one-dimensional.  This results in as much
as a factor of 45 reduction in background variations, after flat field
correction, for a $2048 \times 2048$ pixel CCD.  Because background variations
are often a limiting factor in photometry of faint or low-surface-brightness
objects, even conventional telescopes are sometimes operated in drift-scan mode
in order to take advantage of this and other benefits (\cite{S.92},
\cite{S.94}).

The TDI technique is particularly important for telescopes which have no
tracking capability.  Liquid-mirror telescopes (LMTs) employ rotating primary
mirrors surfaced with a liquid metal such as mercury.  Because the rotation
axis must be vertical, these telescopes are best suited for observations near
the zenith, although it is possible in principle to observe at large zenith
angles using specially-designed correctors (\cite{RM87}, \cite{BMW95}).  By
operating their CCD cameras in TDI mode, such zenith-pointing telescopes can
survey large areas of sky as the Earth rotates.  For example, the UBC-NASA
Multiband Survey (UNMS, \cite{HM98}) employs a $2048 \times 2048$ pixel CCD
camera at the prime focus of the 3-meter LMT of the NASA Orbital Debris
Observatory (\cite{PM97}).  This survey covers 20 square degrees and reaches
$m_{AB} \simeq 20.4$ in a single TDI scan through medium-band filters.

For TDI observations, it is essential that the images of all objects drift at
the same rate, on parallel linear tracks, or image spread will result.  For
this reason it is important that telescopes used for TDI observations be
equipped with correctors that reduce the pincushion or barrel distortion that
is usually present.  For a prime-focus configuration, this normally requires a
corrector having at least four optical elements.

After removal of telescope distortion there remains, however, a more
fundamental difficulty.  With the exception of objects on the celestial
equator, the image tracks are not straight, but concave, in the direction of
the nearest celestial pole.  Furthermore, the linear rate of image motion is a
function of declination, so there is no universal CCD scan rate which will
prevent image spread.  These effects can lead to significant image degradation
for telescopes at moderate latitudes employing large-format CCD cameras.

An elegant solution to this problem has been realized with the Great Circle
Camera (\cite{ZSB96}) which uses a combination of rotation and motion in
declination to trace a great circle on the sky.  As a result, the nonlinear
effects are minimized.  While the Great Circle Camera does reduce
curvature effects, it has a fundamental limitation.  Because of the physical
motion required of the camera, the length of the great circle track is limited.
This restriction is particularly severe for zenith-pointing telescopes which
cannot move to follow a great-circle track.  TDI observations would need to be
interrupted frequently while the camera position is reset to follow different
great-circle arcs.

In this paper we propose a new solution to the problems of star-trail
curvature.  By offsetting and tilting the lenses of a suitably-designed
corrector, one can introduce an asymmetric field distortion which compensates
for the nonlinear motion of the images.  As a result, the images track in
parallel straight lines at a common constant rate with high accuracy.  This
effect can be achieved without introducing any additional optics in the light
path.  Moreover, the same corrector can serve telescopes at different latitudes
just by repositioning the optical elements.

The paper is organized as follows:  In Sec.  2 we analyze the geometry of
star-trail curvature and derive formulae for the resulting image displacements.
In Sec.  3 we describe an example of a prime-focus corrector which compensates
for these effects.  The performance of the corrector is examined in Sec.  4
followed by our conclusions.

\section{Analysis of Star-Trail Curvature}

Star-trail curvature effects have been discussed previously by
(\cite{GH92}, \cite{S.96} and \cite{C98}) who give approximate
formulate for the distortions. Our aim here is to present a simple, exact, 
derivation of the relevant relations as well as useful approximate formulae.
The geometrical picture is illustrated in Figure 1.  The circle represents the
celestial sphere, which for convenience we take to have unit radius.  Its center
is the origin of a three-dimensional Cartesian coordinate system $(x,y,z)$.  The
$z$-axis is aligned with the North Celestial Pole (NCP) hence the $x$-$y$ plane
intersects the sphere at the celestial equator.  For clarity, the $y$ axis,
which is perpendicular to the page, is not shown in the figure.  Now consider a
telescope whose axis lies in the $x$-$z$ plane at an angle $\delta_0$ to the
celestial equator.  $\delta_0$ is therefore the declination of the center of the
telescope field of view.  For a zenith-pointing telescope, $\delta_0$ is equal
to the latitude of the observatory.  In the telescope's focal plane,
perpendicular to the optical axis, we set up a two-dimensional Cartesian
coordinate system $(X,Y)$, oriented so that the $Y$-axis is parallel to the
$y$-axis.  The optical axis passes through the origin of the $X$-$Y$ coordinate
system, so that $X$ and $Y$ give the North and East position of the image with
respect to the field center.  It is convenient to work with the projection of
the focal plane, through the telescope optics, back onto the celestial sphere.
Figure 1 shows the projected focal plane which is tangent to the celestial
sphere at the point where it is intersected by the optical axis.  Since the
sphere has unit radius, the values of $X$ and $Y$ can be converted to physical
distances in the focal plane by multiplying by the telescope's effective focal
length $F$.

Now consider the sidereal motion of a star at arbitrary declination $\delta$. As
the Earth rotates, the line of sight to the star sweeps out a cone, indicated
in Figure 1 by the `*' symbol.  The intersection of this cone with the tangent
plane defines the track of the star's image on the focal plane.  This track is
by definition a conic section, which is an ellipse, parabola or hyperbola
depending on whether the declination $\delta$ is greater than, equal to, or
less than the co-angle $\pi - \delta_0$ (for a zenith telescope, the
co-latitude).

The equation of this track is easily derived. The equation of the cone is
\begin{equation}
	x^2 + y^2 = z^2 \cot^2\delta~, \label{eq:cone}
\end{equation}
while the plane is defined by
\begin{eqnarray}
	x & = & -X\sin\delta_0 + \cos\delta_0 \nonumber \\
	z & = &  X\cos\delta_0 + \sin\delta_0~. \label{eq:plane}
\end{eqnarray}
Eliminating $x$ and $z$, and setting $y = Y$, we obtain
\begin{equation}
	(\sin^2\delta - \cos^2\delta_0)X^2 -2\sin\delta_0\cos\delta_0 X
	+ (Y^2 + 1)\sin^2\delta -\sin^2\delta_0 = 0~.
\end{equation}
The northerly displacement $X$ of the image as a function of East-West 
position $Y$ is given by the solution
\begin{equation}
	X = {\sin\delta_0\cos\delta_0\over \sin^2\delta - \cos^2\delta_0} 
	\Biggl\{1-\Biggl[1 - {(\sin^2\delta - \cos^2\delta_0)([1 + Y^2]
	\sin^2\delta - \sin^2\delta_0) \over\sin^2\delta_0\cos^2\delta_0}
	\Biggr]^{1/2} \Biggr\}~.
\end{equation}
This equation has a simple Taylor-series expansion:
\begin{equation}
	X = \tan(\delta - \delta_0) + {Y^2\over 2\cot\delta} + \cdots~.
	\label{eq:curve}
\end{equation}

From this we see that the local radius of curvature $R$ measured at the
center of the track is
\begin{equation}
	R \equiv \Bigl|{d^2X\over dY^2}\Bigr|^{-1}_{Y = 0} = \cot\delta~
\end{equation}
which in physical units is $F\cot\delta$. Note that this curvature 
depends only on the declination of the object, and is independent of the 
pointing direction or location of the telescope. Equation (\ref{eq:curve})
may be written
\begin{equation}
	X = X_0 + \Delta X~, \label{eq:X}
\end{equation}
where $X_0 = \tan(\delta - \delta_0)$ denotes the $X$-coordinate of the center 
of the track ($Y = 0$), and $\Delta X$ measures the departure of the track 
from linearity, in the North-South direction. Except at very large 
declinations, the radius of curvature of the trails is nearly constant across 
the field and we may use the value at the field center:
\begin{equation}
	\Delta X \simeq {1\over 2} Y^2 \tan\delta_0~. \label{eq:dx}
\end{equation}

Consider now the rate at which the images move. Let $\phi$ be the azimuth angle
of the object in the $(x,y,z)$ coordinate system, referenced to the $x$-axis. 
As the Earth rotates, $\phi$ increases at a constant rate. From the definition 
of $\phi$, 
\begin{equation}
	Y = y = x\tan\phi~.
\end{equation}
Substituting this in Eqn (\ref{eq:cone}) and using Eqn (\ref{eq:plane})
to eliminate $z$, we obtain
\begin{equation}
	Y = {\sin\phi\over \sin\delta_0\tan\delta + \cos\delta_0\cos\phi}~.
	\label{eq:Y}
\end{equation}

From this we see that the rate of image motion is a function both of the 
declination of the object and the declination of the field center. For a given 
azimuth angle $\phi$, objects which cross the field North of the field center 
($X_0 > 0$) will have smaller $Y$ values that those passing south of the field 
center (in the Northern Hemisphere -- the opposite is true in the Southern 
Hemisphere).

Let $Y = Y_0 + \Delta Y$ where $Y_0$ denotes the $Y$-coordinate of a star,
having the same azimuth angle, whose track passes through the field center 
($X_0 = 0$). From Equations (\ref{eq:curve}), (\ref{eq:X}) and
(\ref{eq:Y}) we obtain
\begin{equation}
	\Delta Y = -X_0 Y_0 \Biggl\{{1 + \tan\delta\tan\delta_0 \over \tan\delta +  
		\cot\delta_0\cos\phi}\Biggr\}~. \label{eq:dy0}
\end{equation}
The factor in parenthesis typically varies very slowly across the field and 
may be replaced by its value at the field center. This gives
\begin{equation}
	\Delta Y \simeq -X_0 Y_0 \tan\delta_0~. \label{eq:dy}
	\label{eq:dY}
\end{equation}

\section{Correcting Curvature Effects}

In the previous section, exact and approximate formulae were developed to
describe the distortions produced by star-trail curvature.  For most practical
situations, the approximate formula of Eqns (\ref{eq:dx}) and {\ref{eq:dy})
give more than enough accuracy.  The maximum distortion can be found by setting
$X_0$ and $Y_0$ equal to half the North-South and East-West extent of the
detector, respectively.  The two distortions are closely related.  For a square
detector, the East-West image smear, due to rate variations, is four times as
large as the North-South smear due to curvature.  (Because the distortion
changes sign as the star crosses the field, the East-West image smear is twice
$\Delta Y$.)

As an example, consider observations of a star at $\delta = 45\arcdeg$ using a
CCD camera with a square field of view of $20\arcmin$ ($X_0 = Y_0 = 0.002909$
radians).  Equations (\ref{eq:dx}) and (\ref{eq:dy}) give maximum distortions
of 0.000004231 radians and 0.000008462 radians ($0.873\arcsec$ and
$1.745\arcsec$) respectively for $\Delta X$ and $\Delta Y$.  These are
significant distortions which would cause unacceptable image degradation for TDI
observations

Fortunately, it is possible to largely eliminate these effects for zenith
telescopes, located even at moderately-high latitudes, by means of an
asymmetric corrector.  The idea is to use a combination of decenters and tilts
of the optical elements in order to introduce distortion that closely matches
that described by Eqns (\ref{eq:dx}) and (\ref{eq:dy}), but with opposite sign.
Such a design has been recently developed for the Large Zenith Telescope (LZT),
which employs a 6 meter f/1.5 liquid-mercury primary mirror (\cite{H.98}.  The
corrector was designed by E.H.R.  to specifications provided by P.H.  The
required image quality was a 50\% encircled energy diameter (EED) of
$0.4\arcsec$ or less over a $10.5\arcmin \times 21\arcmin$ field of view (to
match a $2048 \times 4096$ pixel CCD).  The procedure was to first aim for a
corrector with zero distortion.  Global optimizations were employed to keep the
maximum clear aperture relatively small, in order to reduce cost, while meeting
image-quality specifications over the required field of view.  The corrector
was then reoptimized to introduce the required sidereal distortion.  In this
process, curvatures, tilts, decenters and spacings were all allowed to change.
At first, a wedge was allowed for one element, but it was found that this could
be eliminated, allowing the same lenses to be used at different latitudes.
After optimization for the $+49\arcdeg 17\arcmin$ latitude of the LZT, the lens
curvatures and thicknessess were frozen.  In order to illustrate the
flexibility of the corrector, the design was then reoptimized for the
$+32\arcdeg 59\arcmin$ latitude of the NODO LMT, and the equator ($0\arcdeg$), 
allowing only the tilts, decenters and locations of the lenses to change.

The optical configurations, for latitudes $+49\arcdeg 17\arcmin$ and
$+32\arcdeg 59\arcmin$, are illustrated in Figures 2
and 3.  The corrector employs four lenses all fabricated from the same optical
material.  The details of the corrector design are provided in Table 1,
which gives parameters for the two asymmetric configurations and for the 
symmetric case.  Each line of the table
describes an optical surface, in the order in which light reaches it.  Light
enters parallel to the z axis of a Cartesian coordinate system aligned with the
Surface 1 (the parabolic reflector).  Columns (7), (8), (10) and (11)
describe translations and rotations of the coordinate system in which
subsequent surfaces are defined.  Surfaces following such a transformation are
aligned on the local mechanical axis (z-axis) of the new coordinate system.
The new mechanical axis remains in use until changed by another transformation.
In these transformations, the translation is applied before the rotation.

The column headings for the table are as follows: (1) surface number, (2) material
that the light enters when crossing the surface, (3) radius of curvature
of the surface, (4) the aspheric constant $A$ (except for surface 1 which is a 
pure conic section), defined by the equation
\begin{equation}
  \Delta z = {r^2 \over 1 + \left\{1-(1 + K)y^2/R^2\right\}^{1/2}} + A r^4~,
\end{equation}
where $r$ is the distance from the $z$-axis to a point on the surface and 
$\Delta z$ is the $z$-displacement of the surface with respect to a plane,
(5) distance along the $z$-axis from this surface to the next for the $0\arcdeg$ 
configuration, (6) change in the $z$-axis distance required for the 
$+32\arcdeg 59\arcmin$ configuration, (7) displacement of the origin of the 
coordinate system at the surface, (8) rotation angle of the coordinate system 
at the surface. Columns (9), (10) and (11) are the same as columns (6), (7)
and (8) but for the $+49\arcdeg 17\arcmin$ configuration.

\section{Performance of the Corrector}

The corrector provides a 24 arcmin-diameter unvignetted field of view.  The
effective focal length is 10.000 m which gives an image scale of 20.63 arcsec
per mm.  Table 2, gives the 50\%, 80\% and 100\% EEDs for various field angles
and wavelengths.  The median values of the 50\% EED at 400 nm, 500 nm and 600 -
800 nm are 0.353, 0.162 and 0.203 arcsec respectively.  These values are well
below typical ground-based seeing disk diameters at these wavelengths, which
are typically 0.6 arcsec or more at the best astronomical sites, so the
corrector will not appreciably degrade the image.  There is a small, but
non-negligible, focus shift between the three wavelength regions.  Since the
the corrector is intended to be used with common broadband (or narrower)
filters, this is not an issue for image quality.

Table 3 summarizes the distortion characteristics of the corrector.  For the
field angles specificed in Columns (1) and (2), Columns (3) and (4) give the
target and achieved image displacements.  The difference in position, which
corresponds to the distortion error, is listed in Column (5).  It can be seen
from the table that the distortion errors are typically a few tens of
milliarcsec, the maximum error being $0.074\arcsec$.  Since these residuals are
much less than the seeing diameter, the corrector effectively eliminates
star-trail curvature and rate differentials as a source of image degradation.

Considering the 6 meter diameter and fast focal ratio of the primary mirror,
the corrector is very compact having an overall length, from first element to
focal plane, of 50.5 cm.  The largest lens, which has spherical surfaces, has a
diameter of only 34 cm.  This is quite small for a 6-meter f/1.5 telescope,
which helps reduce the cost of the corrector.  The three smaller lenses have
diameters less than 15 cm.  The rear side of each is a low-order aspheric
surface.

\section{Summary and Discussion}

We have presented an analysis of star-trail curvature effects.  If uncorrected,
these effects can result in substantial image degradation in drift-scan images
obtained with large-format CCDs.  A new technique has been described in which
the curvature effects are compensated by means of a corrector lens employing
decentered and tilted elements.  As an example, we have discussed a compensated
prime-focus corrector designed for the 6-meter LZT.

The corrector requires no additional optical components other than the four
elements normally needed to remove telescope aberrations, including distortion.
By varying the positions of the the elements, the corrector can be used with a
zenith-pointing telescope at any latitude up to at least $\pm 50\arcdeg$.  It
provides a $24\arcmin$ unvignetted field with median 50\% EED of $0.2\arcsec$
and a maximum distortion error of order $0.07\arcsec$.

An interesting question is how much image quality is sacrificed in order to
achieve the distortion correction.  The instantaneous image quality of the
$49\deg$ configuration is slightly worse than that of a similar corrector
optimized for zero distortion.  However the difference is much less than the
improvement in integrated image quality provided by the distortion correction.
For example, reoptimizing the corrector for zero-distortion, allowing the lens
shapes to change as well as the separations, results in only a 25\% improvement
in the worst-case RMS image spot diameters.

Can this design be extended to larger fields of view, and how would the result
compare to that of the Great-Circle camera?  While we have not explicitly
investigated this question, during the course of our work an earlier design was
made for a corrector which has a $30\arcmin$ diameter field of
view (to accommodate a $4096 \times 4096$ pixel CCD) and assumes a 5-m f/1.8
primary mirror.  The image quality at $49\arcdeg$ is comparable to that of the
present design.  Because of the smaller primary mirror, a strict comparison
with the present design is not possible, but it does show that wider-field
designs are possible.

In this context, it should be pointed out that even the Great-Circle camera is
not entirely distortion-free.  The image displacements can be obtained from
Equations (\ref{eq:curve}) and (\ref{eq:dy0}) by setting $\delta_0 = 0$ and
taking $\delta$ to be the field angle $X$.  This gives $\Delta X =
Y^2\tan\delta/2 \simeq XY^2/2$ and $\Delta Y = 0$.  So while the great-circle
camera is free from rate-variation, there is a small amount of star-trail
curvature.  This results in an image smear which increases in proportion to the
cube of the field diameter and becomes substantial for field sizes of order one
degree (the image displacement is $0.548\arcsec$ for a $1\arcdeg \times
1\arcdeg$ field).

While the corrector design presented here was specifically developed for a
zenith-pointing telescope, the same optical design could also be used with a
conventional telescope.  Because the curvature effects depend only on
declination, compensation could be introduced for any field center by means of
mechanical actuators that would adjust the decenters and tilts of the
individual elements in the corrector.  The corrector would be reconfigured for
each pointing of the telescope, and then held fixed during the integration.
This would be the case even if the scanning is done at a non-sidereal rate, so
long as the declination of the field center remains constant during the
exposure.

\acknowledgments

We thank the referee, Dr.  Stephen Shectman, for helpful comments.  Research
support for PH is provided by grants from the Natural Sciences and Engineering
Research Council of Canada (NSERC).  The LZT is funded by NSERC Collaborative
Project Grant no.  CPG0163307 and the University of British Columbia.

\clearpage
(\cite{GH92}, \cite{S.96} and \cite{C98}) who obtained approximate

\clearpage
\begin{deluxetable}{llrrrrrrrrr}
\tablenum{1}
\footnotesize
\tablecaption{Corrector Specifications \label{tbl1}}
\tablewidth{0pt}
\tablehead{
&&&& \multicolumn{1}{c}{$0\arcdeg$} & \multicolumn{3}{c}{$+32\arcdeg 59\arcmin$} & 
\multicolumn{3}{c}{$+49\arcdeg 17\arcmin$} \nl
&&&& \hrulefill & \multicolumn{3}{c}{\hrulefill} & \multicolumn{3}{c}{\hrulefill}\nl
\colhead{Surf.} & \colhead{Type} & \colhead{$R$} & \colhead{$A$} 
& \colhead{$t$} & \colhead{$\Delta t$} & \colhead{$XDE$} & \colhead{$BDE$} & \colhead{$\Delta t$} 
& \colhead{$XDE$} & \colhead{$BDE$} \nl
&& (mm) && (mm) & (mm) & (mm) & (\arcdeg) & (mm) & (mm) & (\arcdeg)
} 
\startdata
0  & object &  $\infty$  &   ---      & $\infty$  &  0.000 &   ---    &   ---  & 0.000 &   ---    &   ---  \nl
1  & mirror & -18000.000 & K = -1.000 & -8535.429 &  0.870 &   ---    &   ---  & 0.000 &   ---    &   ---  \nl
2  & BK7    &   -201.785 &   ---      &   -30.000 &  0.000 &  -8.193  &  3.584 & 0.000 & -14.160  &  6.215 \nl
3  & air    &   -200.549 &   ---      &  -253.805 & -0.660 &   ---    &  ---   & 0.000 &   ---    &  ---   \nl
4  & BK7    &   -799.735 &   ---      &   -12.000 &  0.000 & -11.002  & -5.250 & 0.000 & -19.215  & -9.150 \nl
5  & air    &   -105.956 &  0.686e-08 &   -56.097 & -1.083 &   ---    &  ---   & 0.646 &   ---    &  ---   \nl
6  & BK7    &    594.610 &   ---      &   -20.000 &  0.000 &  11.817  &  2.839 & 0.000 &  21.177  &  4.941 \nl
7  & air    &    236.851 & -0.212e-07 &   -40.216 &  0.782 &   ---    &  ---   & 0.980 &   ---    &  ---   \nl
8  & BK7    &   -131.008 &   ---      &   -50.000 &  0.000 & -11.998  &  0.822 & 0.000 & -21.443  &  1.510 \nl
9  & air    &   -540.262 &  0.227e-06 &   -15.193 &  0.260 &   ---    &  ---   & 0.358 &   ---    &  ---   \nl
10 & BK7    & $\infty$   &   ---      &    -7.000 &  0.000 &   0.169  & -1.023 & 0.000 &   0.444  & -1.835 \nl
11 & air    & $\infty$   &   ---      &    -6.000 &  0.000 &   ---    &  ---   & 0.000 &   ---    &  ---   \nl
12 & quartz & $\infty$   &   ---      &    -5.000 &  0.000 &   ---    &  ---   & 0.000 &   ---    &  ---   \nl
13 & air    & $\infty$   &   ---      &   -11.000 &  0.000 &   ---    &  ---   & 0.000 &   ---    &  ---   \nl
14 & image  & $\infty$   &   ---      &     ---   &   ---  &   ---    &  ---   &  ---  &   ---    &  ---   \nl
\enddata
\end{deluxetable}

\begin{deluxetable}{ccccccccccc}
\tablenum{2}
\small
\tablecaption{Encircled Energy Diameters for $+49\arcdeg 17\arcmin$ \label{tbl3}}
\tablewidth{0pt}
\tablehead{
\colhead{$X$} & \colhead{$Y$} & \multicolumn{3}{c}{400 nm} & \multicolumn{3}{c}{500 nm} &
\multicolumn{3}{c}{600 - 800 nm} \nl
&& 50\% & 80\% & 100\% & 50\% & 80\% & 100\% & 50\% & 80\% & 100\% \nl
($\arcdeg$) & ($\arcdeg$) & 
%(mm) & (mm) & (mm) & (mm) & (mm) & (mm) & (mm) & (mm) & (mm)
($\arcsec$) & ($\arcsec$) & ($\arcsec$) & ($\arcsec$) & ($\arcsec$) & ($\arcsec$) & ($\arcsec$) &
($\arcsec$) & ($\arcsec$)
} 
\startdata
%
% these are the values in arcsec
%
-0.088 & 0.000 & 0.353 & 0.502 & 0.698 & 0.162 & 0.265 & 0.351 & 0.172 & 0.244 & 0.649 \nl
 0.000 & 0.000 & 0.154 & 0.252 & 0.524 & 0.072 & 0.107 & 0.132 & 0.176 & 0.259 & 0.676 \nl
 0.088 & 0.000 & 0.211 & 0.514 & 1.046 & 0.134 & 0.179 & 0.360 & 0.199 & 0.346 & 0.857 \nl
-0.088 & 0.088 & 0.371 & 0.509 & 0.780 & 0.153 & 0.291 & 0.576 & 0.203 & 0.320 & 0.918 \nl
 0.000 & 0.088 & 0.220 & 0.313 & 0.853 & 0.097 & 0.154 & 0.357 & 0.158 & 0.264 & 0.714 \nl
 0.088 & 0.088 & 0.238 & 0.631 & 1.453 & 0.162 & 0.222 & 0.584 & 0.218 & 0.351 & 0.787 \nl
-0.088 & 0.176 & 0.396 & 0.568 & 1.297 & 0.196 & 0.357 & 1.107 & 0.285 & 0.495 & 1.184 \nl
 0.000 & 0.176 & 0.370 & 0.569 & 1.826 & 0.353 & 0.533 & 1.478 & 0.250 & 0.417 & 1.019 \nl
 0.088 & 0.176 & 0.370 & 0.569 & 1.826 & 0.353 & 0.533 & 1.478 & 0.397 & 0.661 & 1.380 \nl
%
% the data below are in mm
%
%-0.088 & 0.000 & 0.01710 & 0.02435 & 0.03382 & 0.00784 & 0.01283 & 0.01704 & 0.00836 & 0.01181 & 0.03147 \nl
% 0.000 & 0.000 & 0.00745 & 0.01224 & 0.02540 & 0.00351 & 0.00521 & 0.00640 & 0.00851 & 0.01257 & 0.03277 \nl
% 0.088 & 0.000 & 0.01025 & 0.02490 & 0.05069 & 0.00650 & 0.00869 & 0.01743 & 0.00966 & 0.01678 & 0.04154 \nl
%-0.088 & 0.088 & 0.01797 & 0.02467 & 0.03781 & 0.00741 & 0.01410 & 0.02791 & 0.00985 & 0.01549 & 0.04452 \nl
% 0.000 & 0.088 & 0.01066 & 0.01517 & 0.04134 & 0.00468 & 0.00745 & 0.01733 & 0.00765 & 0.01281 & 0.03461 \nl
% 0.088 & 0.088 & 0.01153 & 0.03058 & 0.07044 & 0.00784 & 0.01074 & 0.02833 & 0.01055 & 0.01702 & 0.03815 \nl
%-0.088 & 0.176 & 0.01919 & 0.02756 & 0.06287 & 0.00948 & 0.01732 & 0.05367 & 0.01383 & 0.02398 & 0.05739 \nl
% 0.000 & 0.176 & 0.01794 & 0.02759 & 0.08855 & 0.01712 & 0.02582 & 0.07164 & 0.01212 & 0.02020 & 0.04940 \nl
% 0.088 & 0.176 & 0.01794 & 0.02759 & 0.08855 & 0.01712 & 0.02582 & 0.07164 & 0.01927 & 0.03205 & 0.06690 \nl
\enddata
\end{deluxetable}

\begin{deluxetable}{cccccccc}
\tablenum{3}
\small
\tablecaption{Image Displacements for $+49\arcdeg 17\arcmin$ \label{tbl4}}
\tablewidth{0pt}
\tablehead{
\colhead{$X$} & \colhead{$Y$} & \colhead{$\lambda$} & \multicolumn{2}{c}{$\Delta X$} & 
\multicolumn{2}{c}{$\Delta Y$} & \colhead{Error} 
\vspace{-10pt} \nl &&&\multicolumn{2}{c}{\hrulefill} & \multicolumn{2}{c}{\hrulefill} \nl \vspace{-2pt}
&&& Target & Actual & Target & Actual  \nl
($\arcdeg$) & ($\arcdeg$) & (nm) & ($\arcsec$) & ($\arcsec$) & ($\arcsec$) & ($\arcsec$) & ($\arcsec$)
} 
\startdata
-0.088 & 0.176 & 400 & 1.131 & 1.162 &  1.130 &  1.123 & 0.032 \nl % Z3W5DYF7F8    Z3W5F7F1
  "    &  "    & 600 & 1.131 & 1.123 &  1.130 &  1.091 & 0.040 \nl % Z2W3DYF7F8    Z2W3F7F1
  "    &  "    & 800 & 1.131 & 1.113 &  1.130 &  1.068 & 0.065 \nl % Z2W1DYF7F8    Z2W1F7F1
 0.000 & 0.176 & 400 & 1.131 & 1.114 &  0.000 & -0.014 & 0.022 \nl % Z3W5DYF8F5    Z3W5F8F2
  "    &  "    & 600 & 1.131 & 1.110 &  0.000 &  0.022 & 0.030 \nl % Z2W3DYF8F5    Z2W3F8F2
  "    &  "    & 800 & 1.131 & 1.109 &  0.000 &  0.032 & 0.037 \nl % Z2W1DYF8F5    Z2W1F8F2
 0.088 & 0.176 & 400 & 1.131 & 1.132 & -1.130 & -1.188 & 0.058 \nl % Z3W5DYF9F8    Z3W5F9F3
  "    &  "    & 600 & 1.131 & 1.157 & -1.130 & -1.103 & 0.038 \nl % Z2W3DYF9F8    Z2W3F9F3
  "    &  "    & 800 & 1.131 & 1.166 & -1.130 & -1.065 & 0.074 \nl % Z2W1DYF9F8    Z2W1F9F3
 0.000 & 0.132 & 400 & 0.636 & 0.614 &  0.000 & -0.035 & 0.041 \nl % Z2W5DYF11F10 
  "    &  "    & 600 & 0.636 & 0.575 &  0.000 & -0.025 & 0.066 \nl % Z2W3DYF11F10 
  "    &  "    & 800 & 0.636 & 0.611 &  0.000 & -0.022 & 0.033 \nl % Z2W1DYF11F10 
-0.044 & 0.088 & 400 & 0.283 & 0.264 &  0.565 &  0.554 & 0.023 \nl % Z3W5DYF4F5   
  "    &  "    & 600 & 0.283 & 0.283 &  0.565 &  0.526 & 0.039 \nl % Z2W3DYF4F5   
  "    &  "    & 800 & 0.283 & 0.240 &  0.565 &  0.513 & 0.066 \nl % Z2W1DYF4F5   
 0.044 & 0.088 & 400 & 0.283 & 0.259 & -0.565 & -0.605 & 0.047 \nl % Z3W5DYF6F5   
  "    &  "    & 600 & 0.283 & 0.259 & -0.565 & -0.569 & 0.024 \nl % Z2W3DYF6F5   
  "    &  "    & 800 & 0.283 & 0.259 & -0.565 & -0.553 & 0.027 \nl % Z2W1DYF6F5   
\enddata
\end{deluxetable}

\figcaption[]{Geometry of star-trail curvature.}
\figcaption[]{The LZT corrector configured for observations at 
	$+49\arcdeg 17\arcmin$ declination}
\figcaption[]{The LZT corrector configured for observations at 
	$+32\arcdeg 59\arcmin$ declination}

\clearpage
\plotone{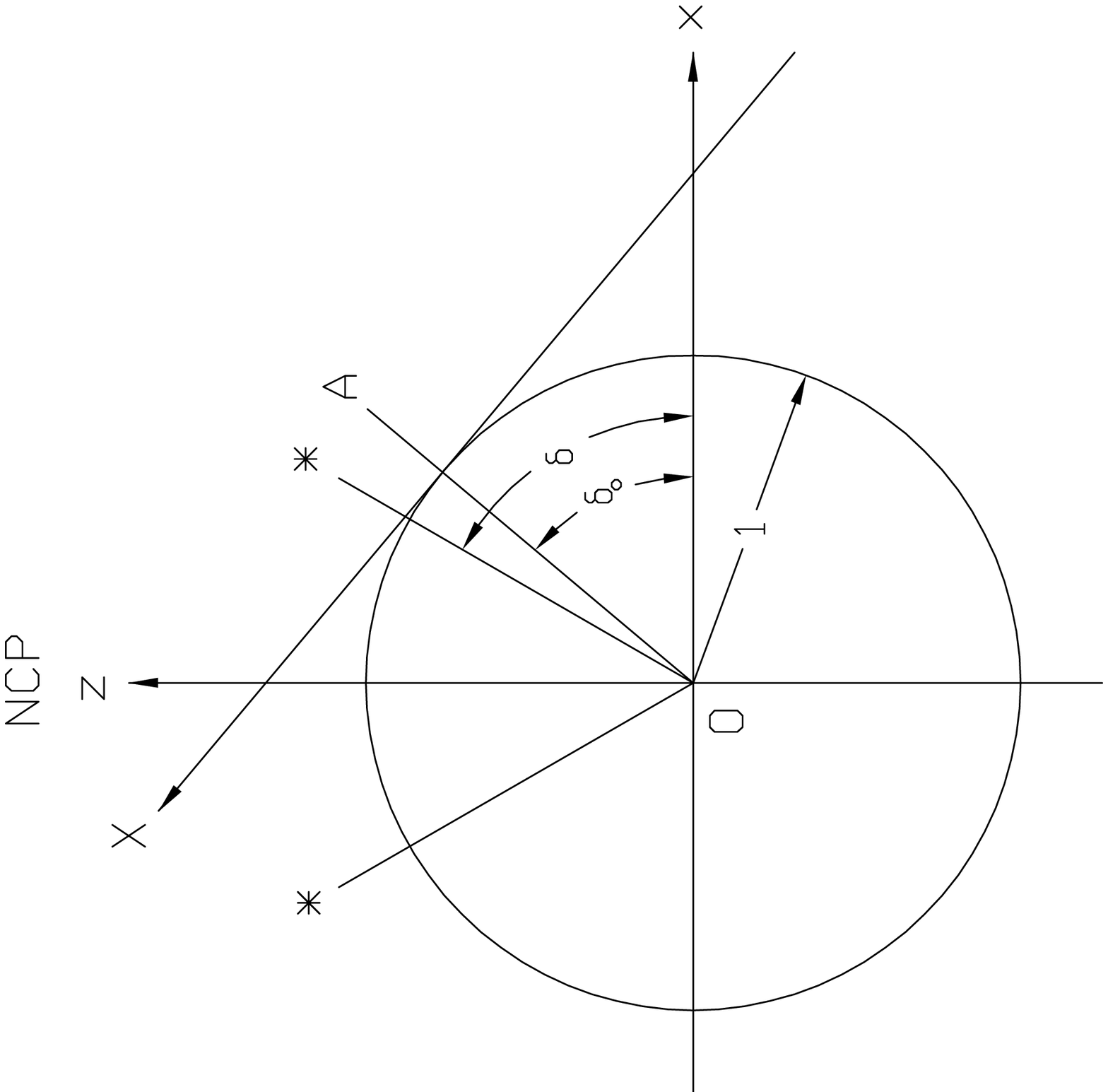}
\clearpage
\plotone{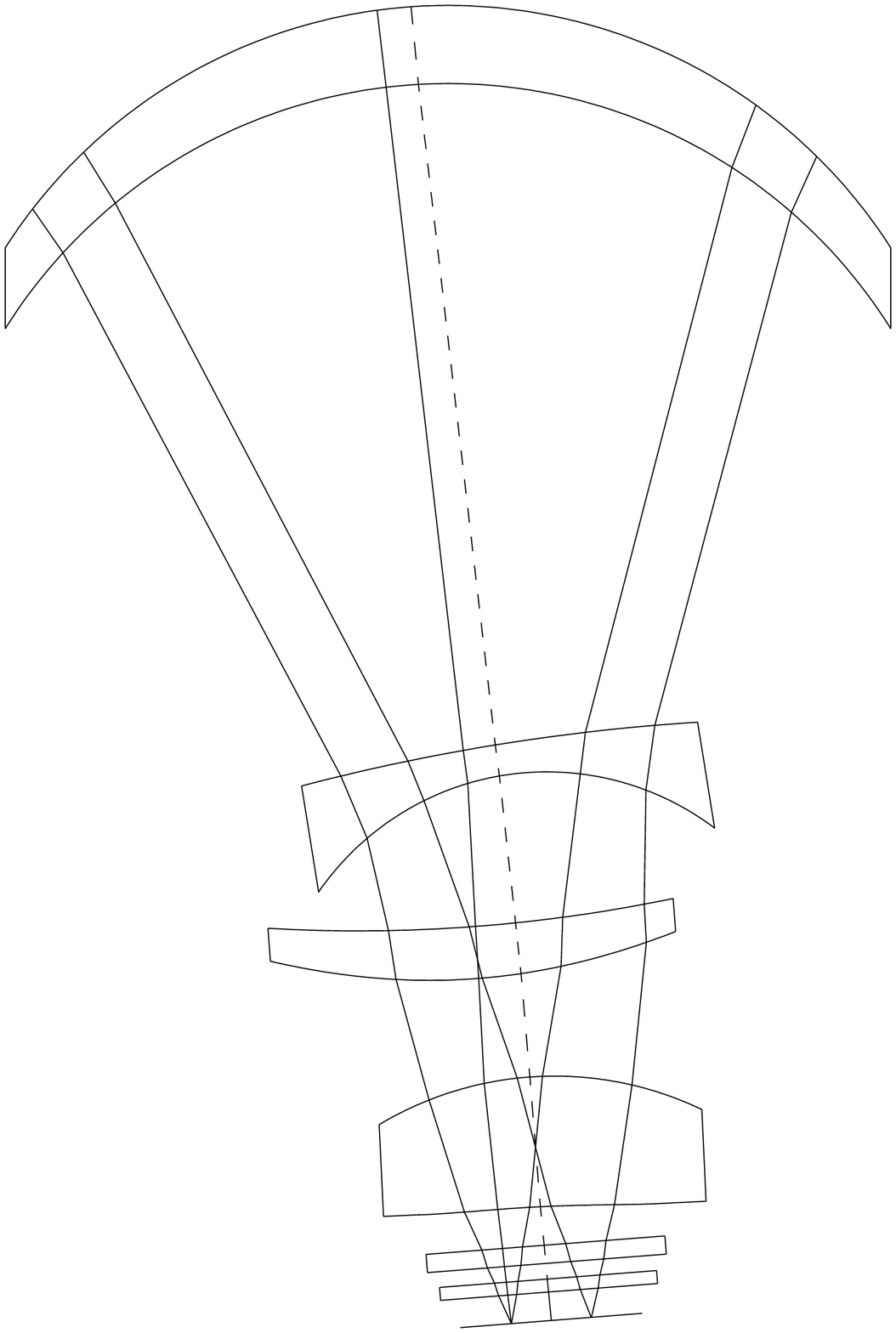}
\clearpage
\plotone{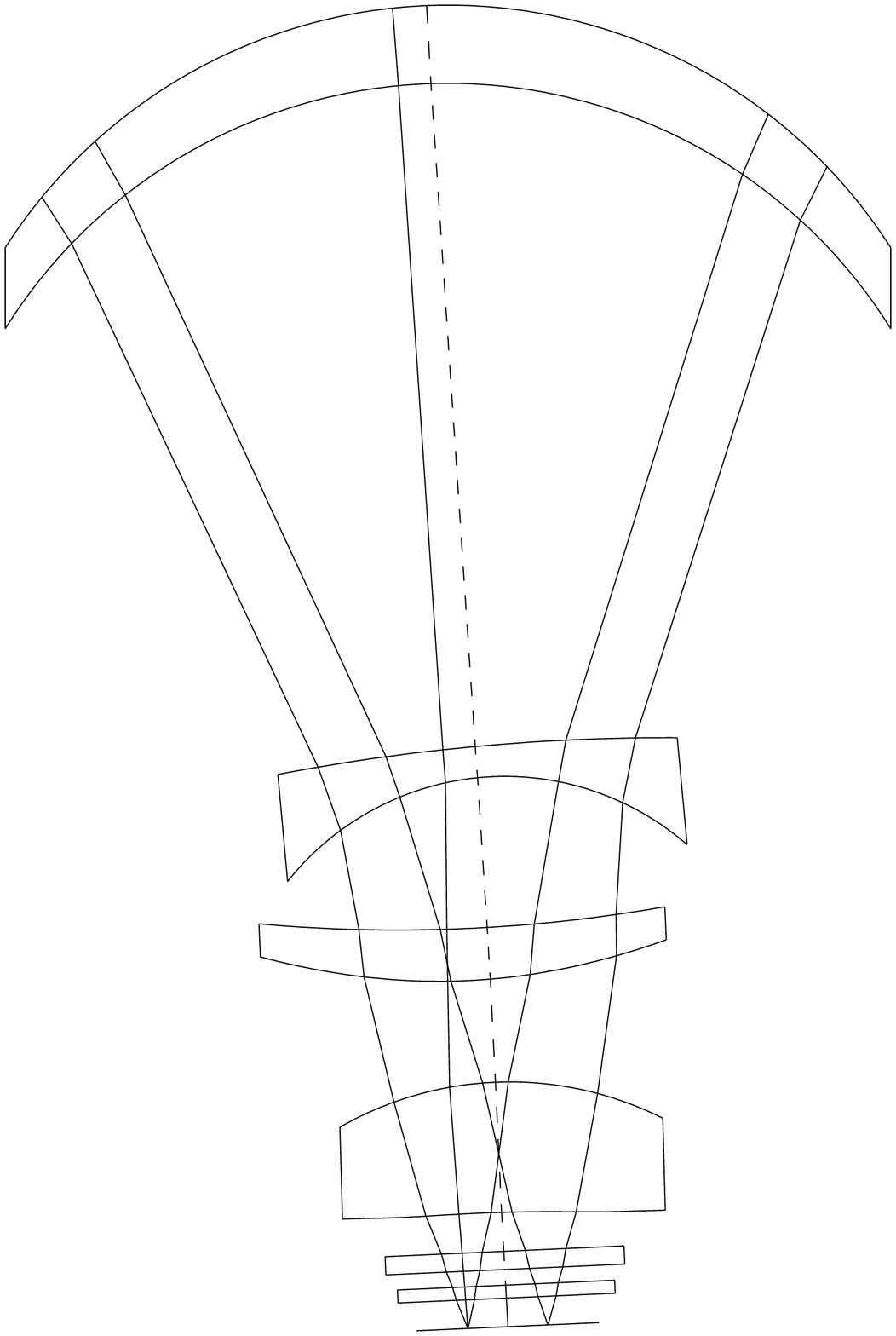}

\end{document}